\documentclass[letter,longauth]{aa}  
\usepackage{graphicx}
\usepackage{txfonts}
\usepackage{xspace}
\usepackage[T1]{fontenc}
\usepackage{footmisc}
\usepackage{soul}

\usepackage{xcolor}

\newcommand{\nnhp}[0]{\chem{N_2H^+}\xspace}
\newcommand{\nnhpoz}[0]{\chem{N_2H^+}{10}\xspace}
\newcommand{\CO}[0]{\chem{^{12}CO}\xspace}
\newcommand{\COoz}[0]{\chem{^{12}CO}{10}\xspace}
\newcommand{\hcn}[0]{\chem{HCN}\xspace}
\newcommand{\hcnoz}[0]{\chem{HCN}{10}\xspace}
\newcommand{\ncrit}[0]{$n_\mathrm{crit}$\xspace}

\usepackage{mychem}
\usepackage{hyperref}
\begin{document}

    \title{Surveying the Whirlpool at Arcseconds with NOEMA (SWAN)}
    \subtitle{I. Mapping the HCN and N$_2$H$^+$ 3mm lines}
\author{Sophia K. Stuber\inst{\ref{i1}}
          \and Jerome Pety\inst{\ref{i2},\ref{i3}}
          \and Eva Schinnerer\inst{\ref{i1}}
          \and Frank Bigiel\inst{\ref{i6}}
          \and Antonio Usero\inst{\ref{i4}}
          \and Ivana Be\v{s}li\'c\inst{\ref{i3}}
          \and Miguel Querejeta\inst{\ref{i4}}
          \and María J. Jiménez-Donaire\inst{\ref{i4},\ref{i4b}}
          \and Adam Leroy\inst{\ref{i7}}
          \and Jakob den Brok\inst{\ref{i5}}
          \and Lukas Neumann\inst{\ref{i6}}
          \and Cosima Eibensteiner\inst{\ref{i6}}
          \and Yu-Hsuan Teng\inst{\ref{i8}}
          \and Ashley Barnes\inst{\ref{i10}}
          \and Mélanie Chevance \inst{\ref{i14},\ref{i15}}
          \and Dario Colombo \inst{\ref{i6}}
          \and Daniel~A.~Dale \inst{\ref{i11}}
          \and Simon C.~O.~Glover\inst{\ref{i14}}
          \and Daizhong Liu\inst{\ref{i13}}
          \and Hsi-An Pan\inst{\ref{i12}}
        }

\institute{
    Max-Planck-Institut für Astronomie, Königstuhl 17, 69117 Heidelberg Germany\label{i1}
    \and 
    IRAM, 300 rue de la Piscine, 38400 Saint Martin d'H\`eres, France\label{i19}\label{i2}
    \and 
    Sorbonne Universit\'e, Observatoire de Paris, Universit\'e PSL, \' Ecole normale sup\`erieure, CNRS, LERMA, F-75005, Paris, France\label{i3}
    \and 
    Argelander-Institut für Astronomie, Universität Bonn, Auf dem Hügel 71, 53121 Bonn, Germany\label{i6}
    \and 
    Observatorio Astron{\'o}mico Nacional (IGN), C/Alfonso XII 3, Madrid E-28014, Spain\label{i4}
    \and 
    Centro de Desarrollos Tecnológicos, Observatorio de Yebes (IGN), 19141 Yebes, Guadalajara, Spain\label{i4b}
    \and 
    Department of Astronomy, The Ohio State University, 140 West 18th Ave, Columbus, OH 43210, USA\label{i7}
    \and 
    Center for Astrophysics | Harvard \& Smithsonian, 60 Garden St., 02138 Cambridge, MA, USA\label{i5}
    \and 
    Center for Astrophysics and Space Sciences, University of California San Diego, 9500 Gilman Drive, La Jolla, CA 92093, USA\label{i8}
    \and 
    European Southern Observatory, Karl-Schwarzschild 2, 85748 Garching bei Muenchen, Germany\label{i10}
    \and 
    Institut  f\"{u}r Theoretische Astrophysik, Zentrum f\"{u}r Astronomie der Universit\"{a}t Heidelberg, Albert- Ueberle-Strasse 2, 69120 Heidelberg, Germany\label{i14}
    \and Cosmic Origins Of Life (COOL) Research DAO, coolresearch.io\label{i15}
    \and 
    Department of Physics \& Astronomy, University of Wyoming, Laramie, WY 82071, USA \label{i11}
    \and 
    Max-Planck-Institut f\"{u}r extraterrestrische Physik, Giessenbachstra{\ss}e 1, D-85748 Garching, Germany\label{i13}
    \and 
    Department of Physics, Tamkang University, No.151, Yingzhuan Road, Tamsui District, New Taipei City 251301, Taiwan\label{i12}
}
   \date{Accepted November 30, 2023}

  \abstract{
  We present the first results from "Surveying the Whirlpool at Arcseconds with NOEMA" (SWAN), an IRAM Northern Extended Millimetre Array (NOEMA)+30m large program that maps emission from several molecular lines at 90 and 110 GHz in the iconic nearby grand-design spiral galaxy M~51 at cloud-scale resolution ($\sim$3\arcsec=125\,pc). 
  As part of this work, we have obtained the first sensitive cloud-scale map of \nnhpoz of the inner $\sim5\,\times 7\,$kpc of a normal star-forming galaxy, which we compare to \hcnoz and \COoz emission to test their ability in tracing dense, star-forming gas. 
  The average \nnhp-to-\hcn line ratio of our total FoV is $0.20\pm0.09$, with strong regional variations of a factor of $\gtrsim 2$ throughout the disk, including the south-western spiral arm and the center. 
  The central $\sim1\,$kpc exhibits elevated \hcn emission compared to \nnhp, probably caused by AGN-driven excitation effects. 
  We find that \hcn and \nnhp are strongly super-linearily correlated in intensity ($\rho_\mathrm{Sp}\sim 0.8$), with an average scatter of $\sim0.14\,$dex over a span of $\gtrsim 1.5\,$dex in intensity.
  When excluding the central region, the data is best described by a power-law of exponent $1.2$, indicating that there is more \nnhp per unit \hcn in brighter regions. 
  Our observations demonstrate that the \hcn-to-CO line ratio is a sensitive tracer of gas density in agreement with findings of recent Galactic studies which utilize \nnhp.
  The peculiar line ratios present near the AGN and the scatter of the power-law fit in the disk suggest that in addition to a first-order correlation with gas density, second-order physics (such as optical depth, gas temperature) or chemistry (abundance variations) are encoded in the \nnhp/\CO, \hcn/\CO and \nnhp/\hcn ratios. 
    }
   \maketitle
%
\section{Introduction}
\label{sec:Introduction}

The emission lines of molecules such as \CO are considered good tracers of the bulk molecular mass distribution (e.g., BIMA-SONG, \citealt{helfer_bima_2003}; PAWS, \citealt{schinnerer_pdbi_2013}; PHANGS, \citealt{leroy_phangsalma_2021}), and are found to correlate with (e.g., infrared) emission tracing recent star formation \citep[e.g.,][]{kennicutt_star_2012, bigiel_star_2008}.  
However, molecular clouds contain a wide range of densities, with star formation typically associated with the densest gas \citep[e.g.,][]{lada_star_2010, lada_star_2012}. 
Extragalactic studies show that CO emission does not distinguish between lower density, bulk molecular gas and the star-forming, dense material with H$_2$ densities of ${\gtrsim} 10^{4}\,\mathrm{cm}^{-3}$ \citep[e.g.,][]{gao_star_2004, jimenez-donaire_empire_2019, querejeta_dense_2019}

Tracers of dense gas are by definition challenging to observe due to the lower abundances of these molecules relative to CO and the smaller volume occupied by the dense phase, which both lead to a much reduced line brightness when compared to CO. 
With typical \hcn-to-CO line ratios in disk galaxies of ${\sim} 1/30$ or lower \citep{usero_variations_2015,bigiel_empire_2016}, extragalactic studies were focusing on \hcn(J=1--0)
finding a tighter correlation between \hcn line emission with star formation rate (SFR) than for CO emission \citep{gao_star_2004, jimenez-donaire_empire_2019}.
The higher critical density (\ncrit) of \hcnoz has led to the common interpretation that this line preferentially traces the denser sub-regions 
of molecular clouds \citep{shirley_critical_2015} making
\hcn a commonly used tracer of dense molecular gas in extragalactic studies \citep{ bigiel_empire_2016,gallagher_spectroscopic_2018, jimenez-donaire_empire_2019, querejeta_dense_2019, beslic_dense_2021, eibensteiner_23_2022, neumann_almond_2023, kaneko_distributions_2023}.

Galactic studies have questioned the use of \hcn as a dense gas tracer at cloud scales (<10\,pc) and favour the use of another molecule, \nnhp, 
which has successfully been detected towards several molecular clouds in the Milky Way \citep[e.g.,][]{pety_anatomy_2017, kauffmann_molecular_2017, barnes_lego_2020, tafalla_characterizing_2021, beuther_cygnus_2022, santa-maria_hcn_2023, tafalla_characterizing_2023}.
Since \nnhp is destroyed in the presence of CO, it is linked to the dense clumps of clouds, where CO freezes to dust grains \citep[e.g.,][]{bergin_cold_2007}. 
This makes \nnhp not only a chemical tracer of cold and dense cores within clouds, but also leads to its emission being beam-diluted and thus even fainter than \hcn (e.g., \nnhp/\CO $\sim 1/100$ at $\sim 150\,$pc scales in a starburst galaxy; \citealt{eibensteiner_23_2022}, \nnhp/\CO $\sim 1/140$ at kpc scales in M~51 \citealt{den_brok_co_2022}).
Extragalactic observations of \nnhp are thus challenging and limited to low-resolution studies \citep[e.g., $\sim$kpc scales;][]{den_brok_co_2022, jimenez-donaire_constant_2023} or individual regions of galaxies \citep[e.g., the center of starburst galaxy NGC~253;][]{martin_alchemi_2021}.  
\citet{jimenez-donaire_constant_2023} summarize Galactic and extragalactic observations of \hcn and \nnhp.
Although challenging, the IRAM Northern Extended Millimetre Array (NOEMA) is capable of obtaining high sensitivity and high angular resolution observations needed to map the distribution of
the faint emission of \nnhp and \hcn in star-forming galaxy disks.

We present the first results from the SWAN (Surveying the Whirpool at Arcsecond with NOEMA) IRAM Large Program (PIs: E. Schinnerer \& F. Bigiel), including the first cloud-scale ($125\,$pc) extragalactic map of \nnhp in the central 5--7\,kpc of the Whirlpool galaxy, a.k.a M~51.
SWAN targets nine molecular lines (\chem{C_2H}{10}, \chem{HNCO}{43}, \hcnoz, \chem{HCO^+}{10}, \chem{HNC}{10}, \nnhpoz, \chem{C^{18}O}{10}, \chem{HNCO}{54}, \chem{^{13}CO}{10}) at $\sim3\,\arcsec$ ($\sim125\,$pc) to study the role of dense gas in the star formation process across galactic environments.

M~51 (NGC~5194) is a nearby \citep[$D = 8.58\,$Mpc;][]{mcquinn_distance_2016} close to face-on \citep[$i = 22\,^{\circ}$, P.A.$ = 173\,^{\circ}$;][]{colombo_pdbi_2014} massive \citep[log$_{10} \, \mathrm{M}_\ast / \mathrm{M}_\odot = 10.5 $;][]{den_brok_co_2022} spiral galaxy 
that hosts a low-luminosity AGN \citep{ho_search_1997, dumas_local_2011, querejeta_agn_2016}.
\hcn has been mapped at 3\arcsec\ resolution (125\,pc) for three circular regions of $\sim3\,$kpc diameter \citep{querejeta_dense_2019} in M~51, at 4\arcsec in the outer spiral arm at $\sim5\,$kpc galactocentric distance \citep{chen_dense_2017} and out to $\sim8\,$kpc in the disk at 1--2\,kpc resolution by EMPIRE \citep{bigiel_empire_2016, jimenez-donaire_empire_2019}. 
\citet{watanabe_spectral_2014} detected both \hcn and \nnhpoz at $\sim$kpc resolution in two 30m pointings in the south-western spiral arm, and  
\citet{den_brok_co_2022} presented \nnhp observations of its center at $\sim$kpc resolution.

Sec.~\ref{sec:Data} describes our observations and data reduction, followed by a comparison of the \nnhp, \hcn, and CO line emission in Sec.~\ref{sec:Results}, a discussion in Sec.~\ref{sec:Discussion} and a summary (Sec.~\ref{sec:Summary}).

\section{Data}
\label{sec:Data}
We use observations from the IRAM large program LP003 (PIs: E. Schinnerer, F. Bigiel) that combines NOEMA (integration time of $\sim214$h) and the 30m single dish observations (about $\sim69$h integration time from EMPIRE, CLAWS, and this program) to map 3-4\,mm emission lines from the central 5$\times$7\,kpc of the nearby galaxy M~51. A detailed description of the observations and data reduction is presented in Appendix~\ref{app:Data}.

The combined data as well as \CO data from PAWS \citep{schinnerer_pdbi_2013} are smoothed to a common angular and spectral resolution of 3\arcsec\ and 10\,km/s per channel. 
We integrate each line by applying the so-called GILDAS-based \textit{island-method} \citep[see][and references therein]{einig23}, where structures with \CO emission above a selected S/N of 2 in the position-position-velocity cube are selected. 
For all lines, the emission is then integrated over the same pixels from the \CO-based 3D-mask.

\section{Results on dense gas in M~51}
\label{sec:Results}
\begin{figure*}[h!]
    \centering
    \includegraphics[width = 0.95\textwidth]{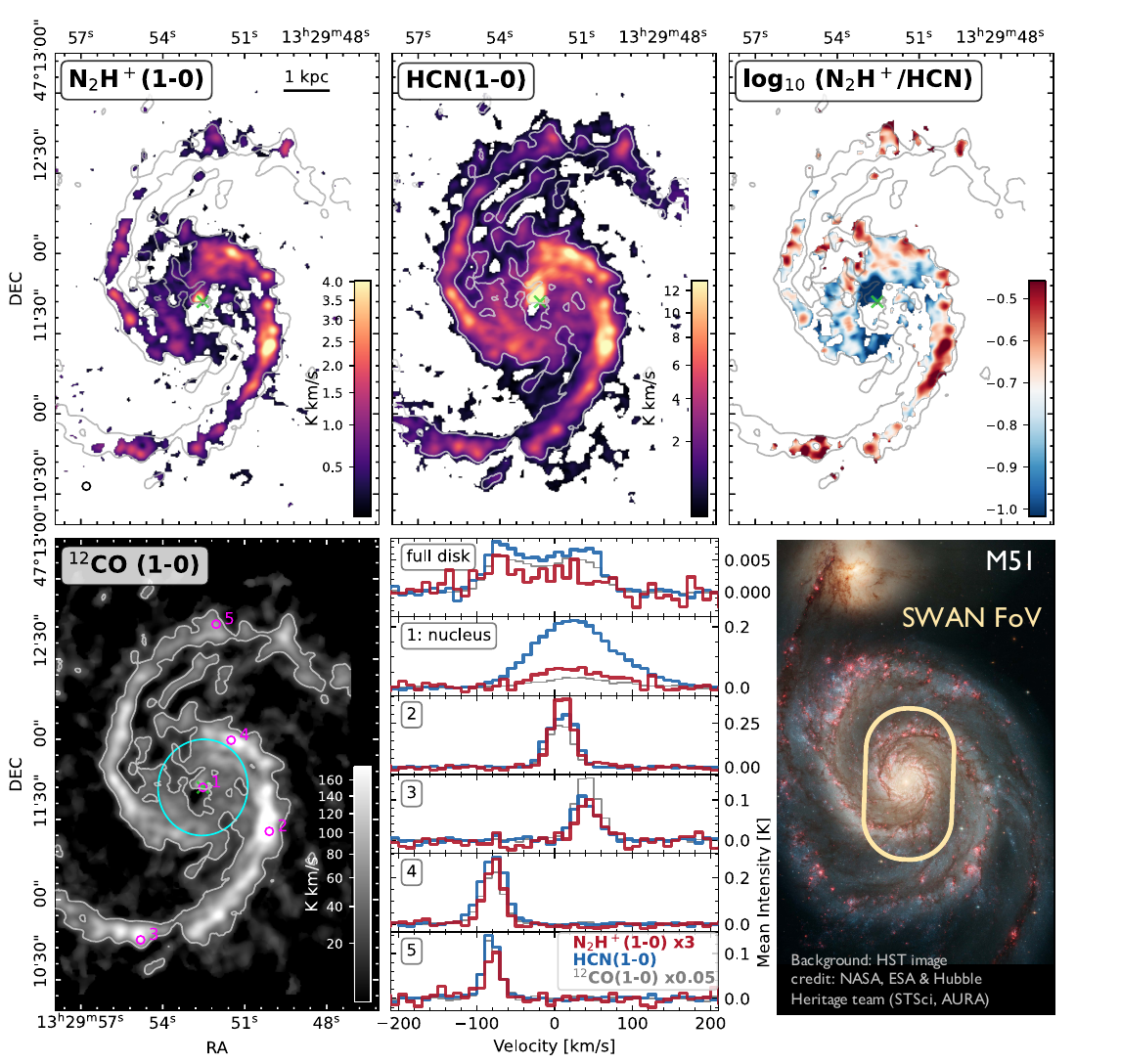}
    \caption{Integrated intensity maps of \nnhp (top left) and \hcn (top center), as well as their ratio (top right) at 3\arcsec ($\sim$ 125\,pc) resolution of the central $5$\,kpc$\,\times\,7$\,kpc in M~51a. 
    The ratio map shows emission above $3\sigma$ for both lines. 
    The beam of $\sim3\arcsec$ is shown in the bottom left corner of the \nnhp map for reference; the location of the galactic center is marked (green $\times$).
    We further display \CO emission at 3$\arcsec$ resolution from the PAWS survey \citep[bottom left; ][]{schinnerer_pdbi_2013} for comparison and show the 30\,K\,km/s contour of \CO for reference in all maps. The central 1.5\,kpc (in diameter) is indicated by a cyan circle in the \CO map.
    Average spectra of five beam-sized regions in the disk (see the \CO map) are shown for \nnhp, \hcn and \CO (bottom center). We scale the spectra by a factor of 3 (\nnhp) and 0.05 (\CO) for easier comparison.  The full-disk spectra contain all pixels in the FoV, shown on top of a HST image (bottom right). 
    }
    \label{fig:Gallery}
\end{figure*}

Our SWAN observations have imaged the line emission of both \hcnoz (hereafter \hcn) and \nnhpoz (hereafter \nnhp) in M~51 at $125\,$pc resolution. 
In order to analyze which physical conditions might impact the brightness of these potential dense molecular gas tracers, we study the \nnhp-to-\hcn ratio across the disk of M~51 (Sec.~\ref{sec:Distribution}), identify regions where the ratio deviates from the global trend (\ref{sec:Lineratios}), and quantify the correlation between \nnhp and \hcn emission (Sec.~\ref{sec:PixelPixelComp}).

\subsection{Distribution of \nnhp and \hcn in the disk of M~51}
\label{sec:Distribution}

Figure~\ref{fig:Gallery} shows the integrated intensity maps of \hcnoz and \nnhpoz, their ratio (upper panels), and the PAWS \CO(1--0) map (bottom left). 
For five beam-sized \nnhp-bright regions in the disk (see \CO map) we extract average spectra of \hcn, \nnhp, and \CO (bottom middle panel). 
\nnhp emission is detected from various extended regions in the disk, including both spiral arms, the molecular ring and interarm regions. 
Both tracers (\nnhp, \hcn) roughly follow the CO brightness distribution with the brightest regions being the galaxy center (denoted as region 1), the southwestern spiral arm (2), as well as the northwestern part of the inner molecular ring (4).

On average, \nnhp is $\sim5$ times fainter than \hcn, and $\sim 80$ times fainter than CO, while line profiles are very similar (FWHM for \nnhp: $\sim 20\,$km/s, \hcn $\sim 30\,$km/s)\footnote{When considering Gaussian line profiles and that \hcn is on average $\sim 5$ times brighter than \nnhp, the inferred linewidths at matched brightness agree for \hcn and \nnhp.}, in agreement with $\sim$kpc observations in NGC~6946 \citep{jimenez-donaire_constant_2023}. 
This remains true, even when imaging our data at 1$\,$km/s spectral resolution. 
As observations of $\sim0.1$\,pc \nnhp clumps in the Milky Way suggest a factor of ${\sim}10$ smaller \nnhp linewidths relative to \hcn  \citep[e.g., 1--2\,km/s;][]{tatematsu_n2h_2008}, our linewidths probably trace cloud-to-cloud velocity dispersion or turbulence scaling with physical lengths. 
Further, the typical \CO luminosity measured per beam
(for details see Appendix~\ref{App:cloudmasses}) indicates multiple clouds per beam. 
We conclude that the similar \hcn and \nnhp linewidths suggest that \hcn and \nnhp spatially coexist inside GMCs at these $>100\,$pc scales and only differ at scales below our resolution.

\subsection{\nnhp-to-\hcn line ratios}
\label{sec:Lineratios}

\begin{table}
\begin{small}
\caption{Typical line ratios of \nnhp, \hcn and \CO in M~51}\label{tab:LineRatios}
\centering
\begin{tabular}{l|l}
\hline\hline
\noalign{\smallskip}
    Line ratio & Average $\pm$ sdev  \\
\noalign{\smallskip}
\hline
\noalign{\smallskip}
    \textbf{N$_2$H$^+$/HCN (1--0)} & 0.20 $\pm$ 0.09\\
    \nnhp/\hcn center & $0.15 \pm 0.05$ \\ 
    \nnhp/\hcn disk  & $0.24 \pm 0.26$  \\
    \hline 
    \noalign{\smallskip}
    \multicolumn{2}{c}{\nnhp/\hcn \textit{in beam-size regions:}}\\
    \noalign{\smallskip}
     region 1 (nucleus) &   $0.08 \pm 0.04 $\\
     region 2 &   $0.33 \pm 0.06 $\\
     region 3 &   $0.28 \pm 0.07 $\\
     region 4 &   $0.23 \pm 0.03 $\\
     region 5 &   $0.24 \pm 0.04 $\\
     \noalign{\smallskip}
     \hline 
     \noalign{\smallskip}
    \textbf{HCN/CO (1--0)} & $0.05 \pm 0.04$  \\
    \hcn/\CO center & $0.10 \pm 0.07$ \\ 
    \hcn/\CO disk & $0.04 \pm 0.02$  \\ 
    \noalign{\smallskip}
    \hline 
    \noalign{\smallskip}
    \textbf{N$_2$H$^+$/CO (1--0)} & $0.012 \pm 0.006$ \\
    \nnhp/\CO center & $0.015 \pm 0.007$  \\ 
    \nnhp/\CO disk & $0.011 \pm 0.005$ \\ 
    \hline 
\end{tabular}
\tablefoot{Average line ratios of \nnhp-to-\hcn as well as \hcn-to-\CO and \nnhp-to-\CO for regions with both \hcn and \nnhp emission $> 3\sigma$ (see Figure~\ref{fig:Gallery}). We list values for the full FoV, as well as for the central 1.5\,kpc in diameter (center) and the remaining disk. For the \nnhp-to-\hcn ratio we further provide ratios of five \nnhp-bright beam-sized regions in the disk selected visually (compare Fig.~\ref{fig:Gallery}). The uncertainty is the standard deviation. }
\end{small}
\end{table}

Our average \nnhp-to-\hcn ratio is $\sim 0.20\pm0.09$ (see Table~\ref{tab:LineRatios}) in regions with detected \nnhp emission ($>3\sigma$) in the integrated intensity map.
Table~\ref{tab:LineRatios} lists average \nnhp/\hcn, \nnhp/\CO and \hcn/\CO ratios derived for the full FoV, the central 1.5\,kpc in diameter, as well as the remaining disk\footnote{The maps are regridded to 1.5\arcsec pixel size to minimize oversampling effects.\label{fnlabel}}. 
The size of the central region is visually set to conservatively encapsulate the area surrounding the center, where low \nnhp-to-\hcn ratios are observed (see Fig.~\ref{fig:Gallery}), but not to include other morphological structures such as the molecular ring at larger radii.  
For these $\gtrsim$kpc regions, the \nnhp-to-\hcn line ratio in the center (disk) is lower (higher) by a factor $\sim 1.3$ ($\sim1.2$) compared to the full FoV value, but still agrees within the uncertainties. 

On $\sim$cloud-size scales ($125$pc), the \nnhp-to-\hcn ratio is significantly ($>3\sigma$) lower in the center (1) than in region 2 in the south-western spiral arm  (see Fig.~\ref{fig:Gallery} \& Table~\ref{tab:LineRatios}) and deviates by a factor of $2.5$ from the full FoV average. 
These findings suggest the presence of systematic trends that drive the high scatter of the full FoV line ratios (see next section).

\subsection{Correlation of \hcn and \nnhp line emission}
\label{sec:PixelPixelComp}

\begin{figure*}[h]
    \centering
    \includegraphics[width = 0.95\textwidth]{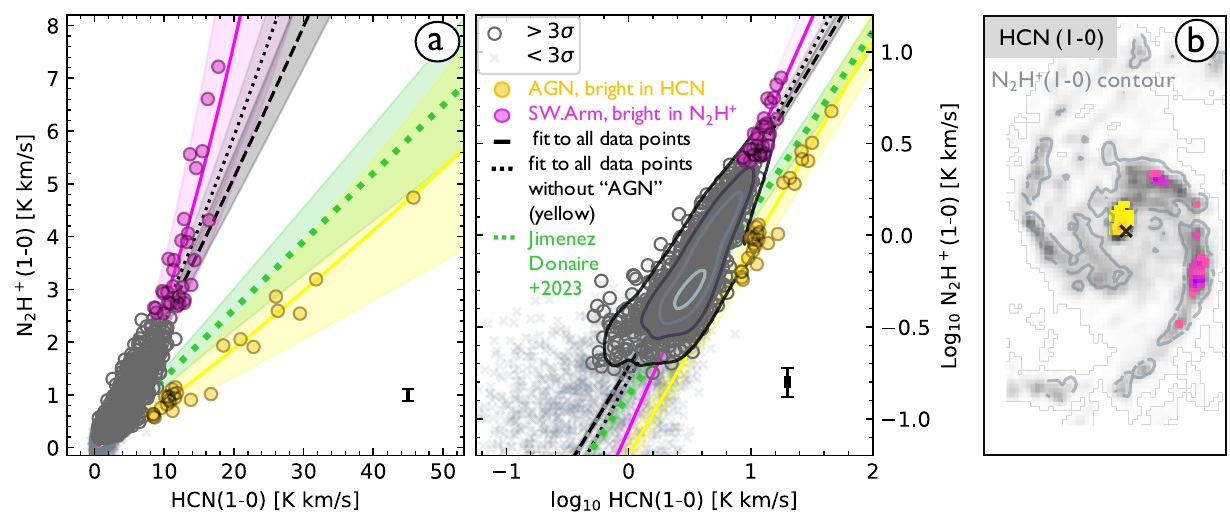}
    \caption{a) Pixel-by-pixel distribution of integrated \nnhp and \hcn emission in linear (left panel) and logarithmic (right panel) scaling. Subsets of pixels are visually isolated based on their high \nnhp (pink, SW.Arm) or \hcn (yellow, AGN) values. Their location relative to the distribution of \hcn emission in M\,51 is shown in b) (contour marks $5\sigma$ \nnhp integrated intensity).
    Power-law fits are applied to all data points (black dashed line), the subsets identified (solid lines), as well as all data points excluding the yellow (AGN affected) subset (black dotted line).
    Fit parameters and Spearman correlation coefficients for all data points and the subsets are given in Table~\ref{tab:Fitparams}.
    Data points below the 3$\sigma$ noise level are presented as grey crosses. 
    The \nnhp-to-\hcn relation from \citet{jimenez-donaire_constant_2023} is shown as a green dotted line. 
    }
    \label{fig:Spikes}
\end{figure*}

\setlength{\tabcolsep}{4pt}
\begin{table}
\caption{Fit parameters and Spearman correlation coefficients of \nnhp as function of \hcn}\label{tab:Fitparams}
\centering
\resizebox{\linewidth}{!}{%
\begin{tabular}{l|llll}
\hline\hline
\noalign{\smallskip}
        Region & Power $a$ & Offset $b$ & $\rho_\mathrm{Sp}$ & $p$-value\\
\noalign{\smallskip}
\hline
\noalign{\smallskip}
        All data & $1.10 \pm 0.02$ & $-0.72 \pm 0.02$ & $ 0.832 \pm 0.009$& $<0.001$\\ 
        All w/o AGN & $1.20 \pm 0.02$ & $-0.79 \pm 0.02$ &$ 0.834 \pm 0.009$ & $<0.001$\\
        AGN, yellow & $1.13 \pm 0.04$ & $-1.20 \pm 0.06$ &  $ 0.90 \pm 0.05$ & $ < 0.001$\\
        SW.Arm, pink & $1.49 \pm 0.09$ & $-1.06 \pm 0.10$ & $ 0.75 \pm 0.05$ &  $ < 0.001$\\ 
\hline        
\end{tabular}
}
\tablefoot{Fit parameters for a linear fit in log-log space (log$_{10}\,I_\mathrm{N_2H^+}= a \cdot \,$log$_{10}\,I_\mathrm{HCN} \,+\,b$), which corresponds to a power-law relation in linear space ($I_\mathrm{N_2H^+}\, = 10^{b} \cdot I_\mathrm{HCN}^{a}$). We add Spearman correlation coefficients $\rho_\mathrm{Sp}$ and corresponding p-values. 
We only consider pixels with significant emission (i.e. $> 3\sigma$).
}
\end{table}

To study how well the \nnhp and \hcn emission correlate, we analyze the pixel-by-pixel distribution of \nnhp intensity as a function of \hcn intensity\footref{fnlabel} (Fig.~\ref{fig:Spikes}, for \nnhp and \hcn as a function of \CO emission see Appendix~\ref{app:Additionalspikes}).

To first order, 
the \nnhp emission is strongly correlated with the \hcn emission (Spearman correlation coefficient $\rho_\mathrm{Sp} = 0.832\pm0.009$, see Table~\ref{tab:Fitparams}).
Though, some data deviate from the correlation. 
The linear presentation (left panel) reveals two clusters with different mean slopes. 
We visually devise subsets of pixels that  
(a) belong to the main cluster containing the bulk of the data points (grey),  
(b) have comparably low \nnhp flux while very high \hcn fluxes (yellow), and
(c) show the highest \nnhp intensities where the apparently linear trend becomes exponential (pink).
Locating these pixels in the \hcn moment-0 map (Fig.~\ref{fig:Spikes} b) reveals that subset (b) originates from the galaxy center (yellow, hereafter "AGN") and subset (c) from the south-western arm (pink, hereafter "SW.Arm"). Pixels in the central part of the galaxy thus follow a distribution that is significantly fainter in \nnhp emission than in \hcn emission compared to the rest of the sample.
We discuss the impact of the AGN in Section~\ref{sec:Discussion:AGN}.

The logarithmic presentation (right panel of Fig.~\ref{fig:Spikes} a) confirms that subset (c, pink) from the south-western arm  follows the bulk data (grey) for a power-law distribution. 
The comparably large scatter in this subset (c, SW.arm) emerges from 
two different spatial locations that have slightly different slopes than the subset's average one (see also Appendix~\ref{App:scatter}).

We fit all data points plus the subsets (emission $>3\sigma$) with linear functions in logarithmic scaling. 
Fit parameters as well as Spearman correlation coefficients ($\rho_\mathrm{Sp}$) and $p$-values are provided in Table~\ref{tab:Fitparams}). 
Details on the fitting process, and the uncertainties derived via jackknifing are given in Appendix~\ref{app:Fitting}.
For all subsets, \nnhp emission is similarly well ($\rho_\mathrm{Sp}>0.75$) and super-linearly (best-fit power $a>1$) correlated with \hcn emission.
However, the fit of the central (yellow) data points significantly deviates from the fit for all data points including and excluding the central ones (black dashed and dotted line, see also Appendix~\ref{app:Fitting}). 
The central subset contributes $\sim 7\%$ of the total \hcn and $\sim 4\%$ of the total \nnhp flux in our FoV, and contains most of the brightest \hcn pixels.  
Although the disk data points (without AGN and SW.arm, grey points), can be best described with a linear relation (power $a=0.97\pm0.013$), the power index monotonically increases when the upper limit of the range in integrated \nnhp emission used to select the fitted point is increased. This is likely due to  the scatter around the power law decreasing at the same time as the data explores a larger part of the power law, increasing the range spanned by the data.

The \nnhp-to-\CO and \hcn-to-\CO distributions (Fig.~\ref{app:Additionalspikes}) behave similarly, as the central data points clearly deviate from the bulk distribution. 
Similar to the \nnhp-to-\hcn distribution, the \nnhp-to-\CO distribution (Fig.~\ref{fig:Additionalspikes}) is best described by a super-linear power-law, with its brightest end being mainly populated by the pixels from subset (c, SW.arm) (Table~\ref{tab:FitparamsAppendix}). 
In contrast, the \hcn-to-\CO distribution is best described by a sub-linear to linear power-law. 
We quantify the scatter of these distributions in Appendix~\ref{App:scatter2} and find that for all distributions, the scatter is of order $\sim0.14\,$dex, while the total range covered by the lines cover $\gtrsim 1.5\,$dex.

\subsection{Density-sensitive line ratio}
\label{App:DG}

\begin{figure}
    \centering
    \includegraphics[width = 0.5\textwidth]{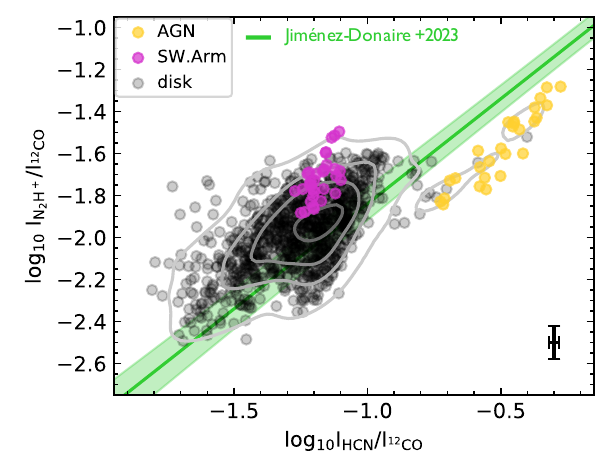}
    \caption{I$_\mathrm{N_2H+}$/I$_\mathrm{CO}$ as function of I$_\mathrm{HCN}$/I$_\mathrm{CO}$ for all data points and subsets above a $3\sigma$ noise level. We show the average uncertainty in the bottom right corner, as well as the best-fit from \citet[][eq.~2]{jimenez-donaire_constant_2023} with a power law slope of $1.0$ derived for extragalactic and Galactic data points (green line). Contours indicate the number density of data points.}
    \label{fig:DG_DG}
\end{figure}

The ratio of emission lines from \hcn and \CO ($f = I_\mathrm{HCN}/ I_\mathrm{CO}$) has been commonly used as an indication of the (average) gas density $f_\mathrm{dense}$ \citep[e.g.,][]{usero_variations_2015, bigiel_empire_2016,jimenez-donaire_empire_2019}. We compare $I_\mathrm{N_2H+}/I_\mathrm{CO}$ to $f$ in Fig.~\ref{fig:DG_DG} and find them correlated ($\rho_\mathrm{Sp}= 0.70$, $p$-value $< 0.001$). 
82\% (97\%) of our data points agree within 3$\sigma$ (5$\sigma$) with the fit from \citet[][eq. 2]{jimenez-donaire_constant_2023} with a power-law index of 1.0 obtained when fitting all available Galactic and extragalactic data (green line). 

Our result indicates that, to first order, both line ratios are correlated. 
Although the difference has a low statistical significance, the AGN subset (b, yellow) is offset from the remaining data, and the \nnhp-bright SW.arm subset (c, pink) clusters at higher \nnhp/\CO values. 

\section{Discussion on molecular gas density in M~51}
\label{sec:Discussion}

We discuss which physical conditions might impact the brightness of the potential dense molecular gas tracers based on the found distribution of \nnhp and \hcn in the disk (Sec.~\ref{sec:Distribution}), the high/low line ratios in isolated regions (Sec.~\ref{sec:Lineratios}), and the correlations between emission of \nnhp and \hcn (Sec.~\ref{sec:N2HP_HCN_relation}).

\subsection{The AGN impacts the central emission in M~51}
\label{sec:Discussion:AGN}

The \nnhp-to-\hcn line ratio is significantly lower in the center of M~51 compared to regions in the disk, and the central data points are offset from those in the disk (Fig.~\ref{fig:Spikes}). 
In contrast, the \hcn-to-\CO 
ratio is higher in the center compared to the remaining disk (compare Fig.~\ref{fig:Additionalspikes}), in agreement with \citet{jimenez-donaire_empire_2019} at $\sim$kpc scales. 
Very high-resolution ($\sim30$\,pc) observations of \hcn and \CO in M~51 by \citet{matsushita_resolving_2015} reveal extraordinarily high \hcn/CO ratios ($>2$) at the location of the AGN, which they explain by infrared pumping, possibly weak \hcn masing and an increased \hcn abundance. 
Electron excitation in the XDR of the AGN might also contribute to the enhanced HCN emission \citep{goldsmith_electron_2017}.
\citet{blanc_spatially_2009} identified [NII]$\lambda6584/$H$\alpha$ line ratios typical of AGN in M~51's central $\sim 700\,$pc spatially coincident with X-ray and radio emission. \hcn and \CO arise both from the outflow associated with nearly coplanar radio jet, with significant  effects seen out to $500\,$pc distance \citep{querejeta_agn_2016}. 
While the central $1.5\,$kpc (diameter) region as used for the line ratios likely overestimates the area of influence of the AGN, our visually selected subset (b) likely underestimates the area impacted. 

While both the \hcn-to-\CO and the \nnhp-to-\CO distribution (Fig.~\ref{fig:Additionalspikes}) show an enhancement in \hcn or \nnhp emission in the central data points, the effect is less strong for \nnhp-to-\CO, as the fit to its central points agrees with the disk fit unlike the \hcn-to-\CO distribution (Appendix~\ref{app:Additionalspikes}). 
This implies that \nnhp is less affected by the AGN than \hcn. 
Galactic studies do not find correlations between \nnhp and mid-infrared photons \citep[e.g.,][]{beuther_cygnus_2022} suggesting that \nnhp is not affected by infrared-pumping via the AGN. 
While an increased temperature in the AGN vicinity can also increase HCN emission \citep{matsushita_resolving_2015, tafalla_characterizing_2023}, this would lead to CO sublimating, reacting with and destroying N$_2$H$^+$ in contrast to our findings. High cosmic ray ionization rates in the AGN surroundings might counter this effect by increasing the N$_2$H$^+$ abundance \citep{santa-maria_submillimeter_2021}, which is not seen for HCN \citep{meijerink_star_2011}.
The complex mechanisms happening in the AGN vicinity will be explored in a future paper.

\subsection{The emerging \nnhp-to-\hcn relation}
\label{sec:N2HP_HCN_relation}

Our global average \nnhp-to-\hcn ratio of $0.20\pm0.09$ agrees well with ratios obtained at $\sim$kpc resolution in M~51 of $\sim 0.14$ for the galaxy center, and $\sim0.19$ in the southern spiral arm \citep[][]{watanabe_spectral_2014,aladro_lambda_2015}.
A recent literature compilation \citep{jimenez-donaire_constant_2023} finds that a \nnhp-to-\hcn ratio of $0.07-0.22$ for extragalactic regions, and $\sim 0.05-0.23$ when including Galactic sources.
Line ratios of five $\sim$kpc size regions in NGC~6946 range between $0.12-0.20$, leading to a global ratio of $0.15\pm0.03$, or a linear fit in log-space of power $0.99\pm0.04$ and offset $0.87\pm0.04$ \citep{jimenez-donaire_constant_2023} shown for references in Fig.~\ref{fig:Spikes}.
This fits agrees with our fit to the central subset (b), but shows a significant ($>3\sigma$) deviation from our fits focusing on the disk.

Given the AGN impact (Section~\ref{sec:Discussion:AGN}), we consider the power-law fit without the central subset being most representative of typical conditions: 
The \nnhp emission as a function of HCN emission in the disk at $125\,$pc can be described as:  
\begin{equation}
    \mathrm{log}_{10} \,I_\mathrm{N_2H^+}  = \left(1.20\pm0.02 \right) \cdot \mathrm{log}_{10}\, I_\mathrm{HCN} - \left(0.825\pm0.009 \right)
\end{equation}

The super-linearity in our relation, driving the discrepancy between our and the literature results, 
comes from the bright south-western spiral arm, where our \nnhp-to-HCN ratio is the highest (Table~\ref{tab:LineRatios}, region 2). 
Strong streaming motions present in the southern spiral arm are likely stabilizing the gas resulting in low star formation efficiencies \citep[][]{meidt_gas_2013}. 
This region (at $\sim 28-38\arcsec$) is at the transition between the normal star formation efficiency and the extremely low star formation efficiency seen further south \citep{querejeta_dense_2019} and has a high dynamical complexity \citep[i.e., coinciding with the co-rotation radius of a m=3 mode;][]{colombo_pdbi_2014}. 
Although its \nnhp-to-\hcn ratio is larger than the global average, it extends the general distribution in a smooth manner (Fig.~\ref{fig:Spikes}), unlike the clearly offset emission from M~51's center.
We speculate the following: 

a) \hcn-bright regions have more dense gas (as traced by \nnhp) than what we would expect from the \hcn intensities. This effect should potentially correlate with resolution, as higher-resolution observations able to resolve clouds would be able to better isolate the spatially smaller dense clumps.

b) Galactic studies find that \hcn luminosity is sensitive to far-UV light from young massive stars \citep[e.g.,][]{pety_anatomy_2017, kauffmann_molecular_2017, santa-maria_hcn_2023}. 
\hcn emission is linked to dense molecular clouds, but also well correlated with regions of recent star formation. 
This effect is not seen for \nnhp, 
which is abundant in cold and dense regions where the depletion of CO onto dust grains inhibits the main route of \nnhp destruction.
In the southern-spiral arm, where star-formation is found to be comparably lower, this could explain our power law of 1.2.

As the focus of this study is the comparison of \hcn to \nnhp emission, we selected pixels in the disk where \nnhp is detected. 
Since \nnhp is a chemical tracer of dense gas, we thus selected regions where dense gas can be expected.  
This can introduce a bias towards higher values, as we potentially mask out regions of low \nnhp emission.

\section{Summary and Conclusion}
\label{sec:Summary}

We present the first map of \nnhp(J=1--0) and \hcnoz from the NOEMA+30m large program SWAN in the central $5\times7$\,kpc of the nearby star-forming disk galaxy M~51 at cloud-scale resolution of 125\,pc (3\arcsec). We study where the chemical dense gas tracer \nnhp emits with respect to larger-scale dynamical features and how it relates to emission from other molecules such as \hcn and CO.
Comparing these lines, we find the following: 
\begin{enumerate}
    \item Extended \nnhp emission is detected from various regions across the disk, with the brightest emission found in the south-western spiral arm, followed by the center and the north-western end of the molecular ring. \hcn emission is bright in the same regions, but it shows the highest intensity in the center. 
    
    \item We find an average \nnhp-to-\hcn ratio of $0.20\pm0.09$ for regions detected in \nnhp emission ($>3\sigma$) with strong variations throughout the disk of up to a factor of $\sim 2-3$ in the south-western spiral arm and the center that hosts an AGN.
    \nnhp and \hcn emission are strongly correlated ($\rho_\mathrm{Sp}\sim0.83$), but the central 1.5\,kpc clearly deviates. 
    The disk emission can be described with a super-linear power-law function of index $1.20\pm0.02$, indicating that \hcn-bright regions have higher gas densities as traced by \nnhp than we would infer from their \hcn emission alone. 

    \item 
    The \nnhp-to-\hcn ratio is significantly lower in the M\,51's center where an AGN is present, and its distribution is offset from the bulk of the disk data. 
    The affected region accounts for $\sim 9\%$ of the total \hcn emission and $\sim 4\%$ of total \nnhp emission in pixels in our FoV where \nnhp is detected. 
    Mid-IR pumping might be one explanation for the bright and enhanced \hcn flux surrounding the AGN. 

\end{enumerate}

\bibliographystyle{aa} 
\bibliography{M51letter.bib} 

\begin{thebibliography}{49}
\expandafter\ifx\csname natexlab\endcsname\relax\def\natexlab#1{#1}\fi

\bibitem[{Aladro {et~al.}(2015)Aladro, Martín, Riquelme, Henkel, Mauersberger, Martín-Pintado, Weiß, Lefevre, Kramer, Requena-Torres, \& Armijos-Abendaño}]{aladro_lambda_2015}
Aladro, R., Martín, S., Riquelme, D., {et~al.} 2015, Astronomy \& Astrophysics, 579, A101, publisher: EDP Sciences

\bibitem[{Barnes {et~al.}(2020)Barnes, Kauffmann, Bigiel, Brinkmann, Colombo, Guzmán, Kim, Szűcs, Wakelam, Aalto, Albertsson, Evans, Glover, Goldsmith, Kramer, Menten, Nishimura, Viti, Watanabe, Weiss, Wienen, Wiesemeyer, \& Wyrowski}]{barnes_lego_2020}
Barnes, A.~T., Kauffmann, J., Bigiel, F., {et~al.} 2020, Monthly Notices of the Royal Astronomical Society, 497, 1972

\bibitem[{Bergin \& Tafalla(2007)}]{bergin_cold_2007}
Bergin, E.~A. \& Tafalla, M. 2007, Annual Review of Astronomy and Astrophysics, 45, 339

\bibitem[{Beuther {et~al.}(2022)Beuther, Wyrowski, Menten, Winters, Suri, Kim, Bouscasse, Gieser, Sawczuck, Christensen, \& Skretas}]{beuther_cygnus_2022}
Beuther, H., Wyrowski, F., Menten, K.~M., {et~al.} 2022, Astronomy \& Astrophysics, 665, A63

\bibitem[{Bešlić {et~al.}(2021)Bešlić, Barnes, Bigiel, Puschnig, Pety, Herrera Contreras, Leroy, Usero, Schinnerer, Meidt, Emsellem, Hughes, Faesi, Kreckel, Belfiore, Chevance, den Brok, Eibensteiner, Glover, Grasha, Jimenez-Donaire, Klessen, Kruijssen, Liu, Pessa, Querejeta, Rosolowsky, Saito, Santoro, Schruba, Sormani, \& Williams}]{beslic_dense_2021}
Bešlić, I., Barnes, A.~T., Bigiel, F., {et~al.} 2021, Monthly Notices of the Royal Astronomical Society, 506, 963

\bibitem[{Bigiel {et~al.}(2008)Bigiel, Leroy, Walter, Brinks, De~Blok, Madore, \& Thornley}]{bigiel_star_2008}
Bigiel, F., Leroy, A., Walter, F., {et~al.} 2008, The Astronomical Journal, 136, 2846

\bibitem[{Bigiel {et~al.}(2016)Bigiel, Leroy, Jiménez-Donaire, Pety, Usero, Cormier, Bolatto, Garcia-Burillo, Colombo, González-García, Hughes, Kepley, Kramer, Sandstrom, Schinnerer, Schruba, Schuster, Tomicic, \& Zschaechner}]{bigiel_empire_2016}
Bigiel, F., Leroy, A.~K., Jiménez-Donaire, M.~J., {et~al.} 2016, The Astrophysical Journal, 822, L26

\bibitem[{Blanc {et~al.}(2009)Blanc, Heiderman, Gebhardt, Evans, \& Adams}]{blanc_spatially_2009}
Blanc, G.~A., Heiderman, A., Gebhardt, K., Evans, N.~J., \& Adams, J. 2009, The Astrophysical Journal, 704, 842

\bibitem[{Chen {et~al.}(2017)Chen, Braine, Gao, Koda, \& Gu}]{chen_dense_2017}
Chen, H., Braine, J., Gao, Y., Koda, J., \& Gu, Q. 2017, The Astrophysical Journal, 836, 101

\bibitem[{Colombo {et~al.}(2014)Colombo, Hughes, Schinnerer, Meidt, Leroy, Pety, Dobbs, García-Burillo, Dumas, Thompson, Schuster, \& Kramer}]{colombo_pdbi_2014}
Colombo, D., Hughes, A., Schinnerer, E., {et~al.} 2014, The Astrophysical Journal, 784, 3

\bibitem[{den Brok {et~al.}(2022)den Brok, Bigiel, Sliwa, Saito, Usero, Schinnerer, Leroy, Jiménez-Donaire, Rosolowsky, Barnes, Puschnig, Pety, Schruba, Bešlić, Cao, Eibensteiner, Glover, Klessen, Diederik~Kruijssen, Meidt, Neumann, Tomičić, Pan, Querejeta, Watkins, Williams, \& Wilner}]{den_brok_co_2022}
den Brok, J.~S., Bigiel, F., Sliwa, K., {et~al.} 2022, Astronomy \& Astrophysics, 662, A89

\bibitem[{Dumas {et~al.}(2011)Dumas, Schinnerer, Tabatabaei, Beck, Velusamy, \& Murphy}]{dumas_local_2011}
Dumas, G., Schinnerer, E., Tabatabaei, F.~S., {et~al.} 2011, The Astronomical Journal, 141, 41, publisher: The American Astronomical Society

\bibitem[{Eibensteiner {et~al.}(2022)Eibensteiner, Barnes, Bigiel, Schinnerer, Liu, Meier, Usero, Leroy, Rosolowsky, Puschnig, Lazar, Pety, Lopez, Emsellem, Bešlić, Querejeta, Murphy, Brok, Schruba, Chevance, Glover, Gao, Grasha, Hassani, Henshaw, Jimenez-Donaire, Klessen, Kruijssen, Pan, Saito, Sormani, Teng, \& Williams}]{eibensteiner_23_2022}
Eibensteiner, C., Barnes, A.~T., Bigiel, F., {et~al.} 2022, Astronomy \& Astrophysics, 659, A173

\bibitem[{{Einig} {et~al.}(2023){Einig}, {Pety}, {Roueff}, {Vandame}, {Chanussot}, {Gerin}, {Orkisz}, {Palud}, {Garcia Santa-Maria}, {de Souza Magalhaes}, {Be{\v{s}}li{\'c}}, {Bardeau}, {Bron}, {Chainais}, {Goicoechea}, {Gratier}, {Guzman Veloso}, {Hughes}, {Kainulainen}, {Languignon}, {Lallement}, {Levrier}, {Lis}, {Liszt}, {Le Bourlot}, {Le Petit}, {Danielsson {\"O}berg}, {Peretto}, {Roueff}, {Sievers}, {Thouvenin}, \& {Tremblin}}]{einig23}
{Einig}, L., {Pety}, J., {Roueff}, A., {et~al.} 2023, arXiv e-prints, arXiv:2307.13009

\bibitem[{Gallagher {et~al.}(2018)Gallagher, Leroy, Bigiel, Cormier, Jiménez-Donaire, Hughes, Pety, Schinnerer, Sun, Usero, Utomo, Bolatto, Chevance, Faesi, Glover, Kepley, Kruijssen, Krumholz, Meidt, Meier, Murphy, Querejeta, Rosolowsky, Saito, \& Schruba}]{gallagher_spectroscopic_2018}
Gallagher, M.~J., Leroy, A.~K., Bigiel, F., {et~al.} 2018, The Astrophysical Journal, 868, L38

\bibitem[{Gao \& Solomon(2004)}]{gao_star_2004}
Gao, Y. \& Solomon, P.~M. 2004, The Astrophysical Journal, 606, 271

\bibitem[{{Gildas Team}(2013)}]{gildas_team_gildas_2013}
{Gildas Team}. 2013, Astrophysics Source Code Library, ascl:1305.010, aDS Bibcode: 2013ascl.soft05010G

\bibitem[{Goldsmith \& Kauffmann(2017)}]{goldsmith_electron_2017}
Goldsmith, P.~F. \& Kauffmann, J. 2017, The Astrophysical Journal, 841, 25

\bibitem[{Helfer {et~al.}(2003)Helfer, Thornley, Regan, Wong, Sheth, Vogel, Blitz, \& Bock}]{helfer_bima_2003}
Helfer, T.~T., Thornley, M.~D., Regan, M.~W., {et~al.} 2003, The Astrophysical Journal Supplement Series, 145, 259

\bibitem[{Ho {et~al.}(1997)Ho, Filippenko, \& Sargent}]{ho_search_1997}
Ho, L.~C., Filippenko, A.~V., \& Sargent, W. L.~W. 1997, The Astrophysical Journal Supplement Series, 112, 315, publisher: IOP Publishing

\bibitem[{Jiménez-Donaire {et~al.}(2019)Jiménez-Donaire, Bigiel, Leroy, Usero, Cormier, Puschnig, Gallagher, Kepley, Bolatto, García-Burillo, Hughes, Kramer, Pety, Schinnerer, Schruba, Schuster, \& Walter}]{jimenez-donaire_empire_2019}
Jiménez-Donaire, M.~J., Bigiel, F., Leroy, A.~K., {et~al.} 2019, The Astrophysical Journal, 880, 127

\bibitem[{Jiménez-Donaire {et~al.}(2023)Jiménez-Donaire, Usero, Bešlić, Tafalla, Chacón-Tanarro, Salomé, Eibensteiner, García-Rodríguez, Hacar, Barnes, Bigiel, Chevance, Colombo, Dale, Davis, Glover, Kauffmann, Klessen, Leroy, Neumann, Pan, Pety, Querejeta, Saito, Schinnerer, Stuber, \& Williams}]{jimenez-donaire_constant_2023}
Jiménez-Donaire, M.~J., Usero, A., Bešlić, I., {et~al.} 2023, Astronomy \& Astrophysics, 676, L11, publisher: EDP Sciences

\bibitem[{Kaneko {et~al.}(2023)Kaneko, Tosaki, Tanaka, \& Miyamoto}]{kaneko_distributions_2023}
Kaneko, H., Tosaki, T., Tanaka, K., \& Miyamoto, Y. 2023, The Astrophysical Journal, 953, 139, aDS Bibcode: 2023ApJ...953..139K

\bibitem[{Kauffmann {et~al.}(2017)Kauffmann, Goldsmith, Melnick, Tolls, Guzman, \& Menten}]{kauffmann_molecular_2017}
Kauffmann, J., Goldsmith, P.~F., Melnick, G., {et~al.} 2017, Astronomy \& Astrophysics, 605, L5, arXiv:1707.05352 [astro-ph]

\bibitem[{Kennicutt \& Evans(2012)}]{kennicutt_star_2012}
Kennicutt, R.~C. \& Evans, N.~J. 2012, Annual Review of Astronomy and Astrophysics, 50, 531

\bibitem[{Lada {et~al.}(2012)Lada, Forbrich, Lombardi, \& Alves}]{lada_star_2012}
Lada, C.~J., Forbrich, J., Lombardi, M., \& Alves, J.~F. 2012, The Astrophysical Journal, 745, 190

\bibitem[{Lada {et~al.}(2010)Lada, Lombardi, \& Alves}]{lada_star_2010}
Lada, C.~J., Lombardi, M., \& Alves, J.~F. 2010, The Astrophysical Journal, 724, 687, publisher: The American Astronomical Society

\bibitem[{Leroy {et~al.}(2021)Leroy, Hughes, Liu, Pety, Rosolowsky, Saito, Schinnerer, Schruba, Usero, Faesi, Herrera, Chevance, Hygate, Kepley, Koch, Querejeta, Sliwa, Will, Wilson, Anand, Barnes, Belfiore, Bešlić, Bigiel, Blanc, Bolatto, Boquien, Cao, Chandar, Chastenet, Chiang, Congiu, Dale, Deger, Brok, Eibensteiner, Emsellem, García-Rodríguez, Glover, Grasha, Groves, Henshaw, Donaire, Kim, Klessen, Kreckel, Kruijssen, Larson, Lee, Mayker, McElroy, Meidt, Mok, Pan, Puschnig, Razza, Sánchez-Bl’azquez, Sandstrom, Santoro, Sardone, Scheuermann, Sun, Thilker, Turner, Ubeda, Utomo, Watkins, \& Williams}]{leroy_phangsalma_2021}
Leroy, A.~K., Hughes, A., Liu, D., {et~al.} 2021, The Astrophysical Journal Supplement Series, 255, 19, publisher: The American Astronomical Society

\bibitem[{Martín {et~al.}(2021)Martín, Mangum, Harada, Costagliola, Sakamoto, Muller, Aladro, Tanaka, Yoshimura, Nakanishi, Herrero-Illana, Mühle, Aalto, Behrens, Colzi, Emig, Fuller, García-Burillo, Greve, Henkel, Holdship, Humire, Hunt, Izumi, Kohno, König, Meier, Nakajima, Nishimura, Padovani, Rivilla, Takano, Werf, Viti, \& Yan}]{martin_alchemi_2021}
Martín, S., Mangum, J.~G., Harada, N., {et~al.} 2021, Astronomy \& Astrophysics, 656, A46, publisher: EDP Sciences

\bibitem[{Matsushita {et~al.}(2015)Matsushita, Trung, Boone, Krips, Lim, \& Muller}]{matsushita_resolving_2015}
Matsushita, S., Trung, D.-V., Boone, F., {et~al.} 2015, The Astrophysical Journal, 799, 26, publisher: The American Astronomical Society

\bibitem[{McQuinn {et~al.}(2016)McQuinn, Skillman, Dolphin, Berg, \& Kennicutt}]{mcquinn_distance_2016}
McQuinn, K. B.~W., Skillman, E.~D., Dolphin, A.~E., Berg, D., \& Kennicutt, R. 2016, The Astrophysical Journal, 826, 21, publisher: The American Astronomical Society

\bibitem[{Meidt {et~al.}(2013)Meidt, Schinnerer, García-Burillo, Hughes, Colombo, Pety, Dobbs, Schuster, Kramer, Leroy, Dumas, \& Thompson}]{meidt_gas_2013}
Meidt, S.~E., Schinnerer, E., García-Burillo, S., {et~al.} 2013, The Astrophysical Journal, 779, 45

\bibitem[{Meijerink {et~al.}(2011)Meijerink, Spaans, Loenen, \& Werf}]{meijerink_star_2011}
Meijerink, R., Spaans, M., Loenen, A.~F., \& Werf, P. P. v.~d. 2011, Astronomy \& Astrophysics, 525, A119, publisher: EDP Sciences

\bibitem[{Neumann {et~al.}(2023)Neumann, Gallagher, Bigiel, Leroy, Barnes, Usero, den Brok, Belfiore, Bešlić, Cao, Chevance, Dale, Eibensteiner, Glover, Grasha, Henshaw, Jiménez-Donaire, Klessen, Kruijssen, Liu, Meidt, Pety, Puschnig, Querejeta, Rosolowsky, Schinnerer, Schruba, Sormani, Sun, Teng, \& Williams}]{neumann_almond_2023}
Neumann, L., Gallagher, M.~J., Bigiel, F., {et~al.} 2023, Monthly Notices of the Royal Astronomical Society, 521, 3348

\bibitem[{Pety {et~al.}(2017)Pety, Guzmán, Orkisz, Liszt, Gerin, Bron, Bardeau, Goicoechea, Gratier, Petit, Levrier, Öberg, Roueff, \& Sievers}]{pety_anatomy_2017}
Pety, J., Guzmán, V.~V., Orkisz, J.~H., {et~al.} 2017, Astronomy \& Astrophysics, 599, A98

\bibitem[{{Pety} \& {Rodr{\'\i}guez-Fern{\'a}ndez}(2010)}]{pety_2010}
{Pety}, J. \& {Rodr{\'\i}guez-Fern{\'a}ndez}, N. 2010, \aap, 517, A12

\bibitem[{Querejeta {et~al.}(2016)Querejeta, Schinnerer, García-Burillo, Bigiel, Blanc, Colombo, Hughes, Kreckel, Leroy, Meidt, Meier, Pety, \& Sliwa}]{querejeta_agn_2016}
Querejeta, M., Schinnerer, E., García-Burillo, S., {et~al.} 2016, Astronomy \& Astrophysics, 593, A118, arXiv: 1607.00010

\bibitem[{Querejeta {et~al.}(2019)Querejeta, Schinnerer, Schruba, Murphy, Meidt, Usero, Leroy, Pety, Bigiel, Chevance, Faesi, Gallagher, García-Burillo, Glover, Hygate, Jiménez-Donaire, Kruijssen, Momjian, Rosolowsky, \& Utomo}]{querejeta_dense_2019}
Querejeta, M., Schinnerer, E., Schruba, A., {et~al.} 2019, Astronomy \& Astrophysics, 625, A19

\bibitem[{Santa-Maria {et~al.}(2021)Santa-Maria, Goicoechea, Etxaluze, Cernicharo, \& Cuadrado}]{santa-maria_submillimeter_2021}
Santa-Maria, M.~G., Goicoechea, J.~R., Etxaluze, M., Cernicharo, J., \& Cuadrado, S. 2021, Astronomy \& Astrophysics, 649, A32

\bibitem[{Santa-Maria {et~al.}(2023)Santa-Maria, Goicoechea, Pety, Gerin, Orkisz, Le~Petit, Einig, Palud, de~Souza~Magalhaes, Bešlić, Segal, Bardeau, Bron, Chainais, Chanussot, Gratier, Guzmán, Hughes, Languignon, Levrier, Lis, Liszt, Le~Bourlot, Oya, Öberg, Peretto, Roueff, Roueff, Sievers, Thouvenin, \& Yamamoto}]{santa-maria_hcn_2023}
Santa-Maria, M.~G., Goicoechea, J.~R., Pety, J., {et~al.} 2023, publication Title: arXiv e-prints ADS Bibcode: 2023arXiv230903186S

\bibitem[{Schinnerer {et~al.}(2013)Schinnerer, Meidt, Pety, Hughes, Colombo, García-Burillo, Schuster, Dumas, Dobbs, Leroy, Kramer, Thompson, \& Regan}]{schinnerer_pdbi_2013}
Schinnerer, E., Meidt, S.~E., Pety, J., {et~al.} 2013, The Astrophysical Journal, 779, 42

\bibitem[{Shetty {et~al.}(2007)Shetty, Vogel, Ostriker, \& Teuben}]{shetty_kinematics_2007}
Shetty, R., Vogel, S.~N., Ostriker, E.~C., \& Teuben, P.~J. 2007, The Astrophysical Journal, 665, 1138, publisher: IOP Publishing

\bibitem[{Shirley(2015)}]{shirley_critical_2015}
Shirley, Y.~L. 2015, Publications of the Astronomical Society of the Pacific, 127, 299

\bibitem[{Tafalla {et~al.}(2021)Tafalla, Usero, \& Hacar}]{tafalla_characterizing_2021}
Tafalla, M., Usero, A., \& Hacar, A. 2021, Astronomy \& Astrophysics, 646, A97, arXiv:2101.02710 [astro-ph]

\bibitem[{Tafalla {et~al.}(2023)Tafalla, Usero, \& Hacar}]{tafalla_characterizing_2023}
Tafalla, M., Usero, A., \& Hacar, A. 2023, arXiv:2309.14414 [astro-ph]

\bibitem[{Tatematsu {et~al.}(2008)Tatematsu, Kandori, Umemoto, \& Sekimoto}]{tatematsu_n2h_2008}
Tatematsu, K., Kandori, R., Umemoto, T., \& Sekimoto, Y. 2008, Publications of the Astronomical Society of Japan, 60, 407, arXiv:0804.0111 [astro-ph]

\bibitem[{Usero {et~al.}(2015)Usero, Leroy, Walter, Schruba, García-Burillo, Sandstrom, Bigiel, Brinks, Kramer, Rosolowsky, Schuster, \& Blok}]{usero_variations_2015}
Usero, A., Leroy, A.~K., Walter, F., {et~al.} 2015, The Astronomical Journal, 150, 115

\bibitem[{Virtanen {et~al.}(2020)Virtanen, Gommers, Oliphant, Haberland, Reddy, Cournapeau, Burovski, Peterson, Weckesser, Bright, van~der Walt, Brett, Wilson, Millman, Mayorov, Nelson, Jones, Kern, Larson, Carey, Polat, Feng, Moore, VanderPlas, Laxalde, Perktold, Cimrman, Henriksen, Quintero, Harris, Archibald, Ribeiro, Pedregosa, \& van Mulbregt}]{virtanen_scipy_2020}
Virtanen, P., Gommers, R., Oliphant, T.~E., {et~al.} 2020, Nature Methods, 17, 261, number: 3 Publisher: Nature Publishing Group

\bibitem[{Watanabe {et~al.}(2014)Watanabe, Sakai, Sorai, \& Yamamoto}]{watanabe_spectral_2014}
Watanabe, Y., Sakai, N., Sorai, K., \& Yamamoto, S. 2014, The Astrophysical Journal, 788, 4, publisher: The American Astronomical Society

\end{thebibliography}

\newpage 
\begin{acknowledgements}
This work was carried out as part of the PHANGS collaboration.
We thank the anonymous referee for their constructive feedback.
This work is based on data obtained by PIs E.Schinnerer and F.Bigiel with the IRAM-30m telescope and NOEMA observatory under project ID M19AA.
SKS acknowledges financial support from the German Research Foundation (DFG) via Sino-German research grant SCHI 536/11-1.
JP acknowledges support from the French Agence Nationale de la Recherche through the DAOISM grant ANR-21-CE31-0010 and from the Programme National “Physique et Chimie du Milieu Interstellaire” (PCMI) of CNRS/INSU with INC/INP co-funded by CEA and CNES.
ES acknowledges funding from the European Research Council (ERC) under the European Union’s Horizon 2020 research and innovation programme (grant agreement No. 694343).
AU acknowledges support from the Spanish grant PID2019-108765GB-I00, funded by MCIN/AEI/10.13039/501100011033. 
MQ and MJJD acknowledge support from the Spanish grant PID2019-106027GA-C44, funded by MCIN/AEI/10.13039/501100011033.
JdB acknowledges support from the Smithsonian Institution as a Submillimeter Array (SMA) Fellow. 
LN acknowledges funding from the Deutsche Forschungsgemeinschaft (DFG, German Research Foundation) - 516405419.
The work of AKL is partially supported by the National Science Foundation under Grants No. 1615105, 1615109, and 1653300.
CE acknowledges funding from the Deutsche Forschungsgemeinschaft (DFG) Sachbeihilfe, grant number BI1546/3-1.
Y-HT acknowledges funding support from NRAO Student Observing Support Grant SOSPADA-012 and from the National Science Foundation (NSF) under grant No. 2108081.
MC gratefully acknowledges funding from the Deutsche Forschungsgemeinschaft (DFG) through an Emmy Noether Research Group, grant number CH2137/1-1. COOL Research DAO is a Decentralized Autonomous Organization supporting research in astrophysics aimed at uncovering our cosmic origins.
SCOG acknowledges support from the DFG via SFB 881 “The Milky Way System” (sub-projects B1, B2 and B8) and from the Heidelberg cluster of excellence EXC 2181-390900948 “STRUCTURES: A unifying approach to emergent phenomena in the physical world, mathematics, and complex data”, funded by the German Excellence Strategy.
HAP acknowledges support by the National Science and Technology Council of Taiwan under grant 110-2112-M-032-020-MY3.

\end{acknowledgements}

\begin{appendix}

\section{Data}
\label{app:Data}

We utilize observations from the IRAM large program LP003 (PIs: E. Schinnerer, F. Bigiel) that used the Northern Extended Millimetre Array (NOEMA) and 30m single dish to map the 4-3\,mm line emission from the central 5$\times$7\,kpc of the nearby galaxy M~51. The observations, data calibration and imaging
resulting in our final datasets are briefly described below. 

\subsection{NOEMA data}

NOEMA observations were taken with 9-11 antennas between January 2020 and December 2021, resulting in a total of 214 hours under average to excellent observing conditions with an average water vapor of $\sim 4\,$mm split across
C (59\%; 126h) and D (41\%; 88h) configuration. Observations of the phase and amplitude calibrators (J1259+516 and J1332+473, replaced by 1418+546 if one of the two was not available) were executed every $\sim 20\,$minutes. 
The mosaic consists of 17 pointings in a hexagonal grid. 
Data reduction was carried out using the IRAM standard calibration pipeline in GILDAS \citep{gildas_team_gildas_2013}. Average to excellent temporal subsets of the observed data were selected based on the relative seeing of the different tracks. Absolute flux calibration was done using IRAM models for MWC349 and LkHa101 providing about 50 independent measurements of the flux of J1259+516 and J1332+473 over a period of about 1 year. This allowed us to confirm that the time variations of the flux of these quasars were relatively smooth.


We detect emission from 9 molecular lines between $\sim$80 and 110 GHz, including HCN(1--0; 88.6 GHz) and \nnhp(1--0; 93 GHz) presented in this paper, but also C$_2$H(1--0), HNCO(4--3), HCO$^+$(1--0), HNC(1--0), C$^{18}$O(1--0), HNCO(5--4) and $^{13}$CO(1--0).
Continuum was subtracted from the $uv$ visibilities by fitting a baseline of order 0 for each visibility, excluding channels in a velocity range of 300\,km/s around the redshifted-frequency of each line. 
For each line, the NOEMA data were resampled to a spectral axis with 10\,km/s resolution, relative to the systemic velocity of v$_\mathrm{sys} = 471.7\,$km/s \citep{shetty_kinematics_2007}. 

\subsection{IRAM-30m single-dish data}

In order to sample all the spatial scales, we need to combine the NOEMA interferometric imaging with single dish data of the HCN\,(1--0) and \nnhpoz emission lines from 
both archival and new observations.

The HCN\,(1--0) emission line was observed as part of the IRAM-30m survey EMPIRE \citep{jimenez-donaire_empire_2019}, using the 3\,mm band (E090) of the dual-polarization EMIR receiver (Carter et al. 2012). \nnhpoz was observed by the IRAM-30m CLAWS survey (055-17, PI: K. Sliwa; \citealt{den_brok_co_2022}), where EMIR was also used to map the 1\,mm (220\,GHz) and 3\,mm (100\,GHz) emission lines in M\,51. In both surveys, the integration time was spread over the full extent of M~51, while the required field-of-view (FoV) for the 30m imaging is the interferometric FoV plus a guard-band to avoid edge effects amounting to 5 square arcmin. We estimate that the EMPIRE and CLAWS projects obtained each about 14 hours of 30m integration time over the relevant field of view. As a rule of thumb to achieve an optimum combination one needs to observe with the 30m telescope the same amount of time as is spent in the compact D
configuration\footnote{See the IRAM technical memo 
IRAM-2008-2, https://cloud.iram.fr/index.php/s/Ney5P2BeN7DAEWX}. Hence we obtained 55 hours of additional IRAM 30m observations (project 19-238 observed in February and April 2020) with a similar tuning as the NOEMA one. In all three cases, we used the on-the-fly/\linebreak[0]{}position switching (OTF-PSW) mode, with emission-free reference positions close to the galaxy. The fast Fourier transform spectrometers (FTS) were used to record the data. We refer to \citet{jimenez-donaire_empire_2019} and \citet{den_brok_co_2022} for details of the observations.

The data were (re)-reduced (1) to ensure a homogeneous treatment, and (2) to avoid unnecessary spatial or spectral regridding. In short, for each observed spectrum, we first extracted a frequency range of 300\,MHz centered on each target line. We then
converted the temperature scale from $T_A^{\star}$ to $T_\mathrm{mb}$ by applying the relevant Ruze formula with the \texttt{CLASS} command \texttt{MODIFY BEAM\_EFF /RUZE}. We computed the velocity scale corresponding to each line's redshifted velocity, and we reprojected the spatial offsets of each observed spectrum to the NOEMA projection center of RA=13:29:52.532, Dec=47:11:41.982. We also subtracted an order 1 polynomial baseline fitted excluding a velocity range of $[-170,+170\, {\rm km/s}]$. We finally gridded all the data on the same spatial and spectral grid as the NOEMA data. The achieved noise levels are 2.5\,K at $29.3''$ and 2.4\,K at $27.9''$ for the \hcn and \nnhp{} (1--0) lines, respectively.

\subsection{NOEMA+30m imaging}

The 30m data are then merged with the NOEMA data in the $uv$ plane using the \texttt{GILDAS} \texttt{UV\_SHORT} command \citep[see][for details]{pety_2010}. The combined data are imaged with \texttt{UV\_MAP} on a grid of $768\times1024$ pixels of 0.31\arcsec size. Högbom-cleaning without cleaning mask was run in order to achieve residuals consistent with a Gaussian distribution of the noise. In practice, we ran it until a stable number of clean components, which depends on the line, was reached. The intensity scale was finally converted from Jy/beam to K. 
The resulting dataset has a rms of $\sim 20\,$mK per 10\,km/s channel at a nominal resolution of $2.1 \times 2.4\arcsec$ for the brightest line, $^{13}$CO(1--0).

We created the following sets of integrated moment-0 maps for the NOEMA+30m \hcn and \nnhp datacubes as well as for the $^{12}$CO(J = 1--0) data from PAWS \citep{schinnerer_pdbi_2013}. All data are convolved to a common angular and spectral resolution of 3\arcsec\ (125\,pc) and 10\,km/s per channel.
We integrated each line by applying the so-called \textit{island-method} based on \CO emission \citep[see][and reference therein]{einig23}. This method isolates connected structures with \CO emission above our selected S/N threshold of 2 in the position-position-velocity ($ppv$) cube and integrates the emission of selected lines over the identified structures along the velocity axis. 
To avoid misleading oversampling effects, we regridded our maps to pixels with sizes of half the beam major axis for all calculations (i.e. 1.5\arcsec). 
Using \CO emission to detect the \textit{islands} ensures that for all lines the same pixels in the $ppv$ cube are used for integration. 
Since \CO is brighter than the other lines, there are more pixels above a given S/N threshold than in \nnhp and \hcn. 
Therefore, 
this can result in otherwise too faint emission being stacked in pixels that would not be selected for integration based on, e.g., \nnhp emission. 

For comparison, we also integrate each line individually, by selecting \textit{islands} based on each line's intensity. This reduces the noise for each line individually while conserving faint emission from connected structures. 
However, this also introduce some bias, as emission is integrated over a varying amount of pixels in different structures for each line. 
While some values slightly change due to these effects, we confirmed that all trends and conclusions remain unchanged.

Pixels where the \nnhp{} integrated emission is significantly (>$3\sigma$) detected, contain $\sim50\%$ of significant \CO flux in our FoV and $\sim 70\%$ of significant \hcn emission in the FoV. 
Pixels where \hcn is significantly detected contain $\sim 90\%$ of the significant \CO emission in our FoV. 
Since the our analysis in section 3 is based on regions where \nnhp is detected, we limit our analysis to a smaller area in the FoV.

\section{Typical cloud mass per beam}
\label{App:cloudmasses}

As \nnhp is abundant in small dense cores, we generally expect the N2H+ emission to arise from regions 
with sizes much smaller than our resolution of 125\,pc, which is slightly worse than the size of massive GMCs. 
Thus, the pixel-by-pixel variations in our observations reflect average physical trends affecting ensembles of multiple clumps.
Our finding in Sec.~\ref{sec:Distribution}, that linewidths of \nnhp are comparable to \hcn in selected beam-size regions, suggests that at our 125\,pc scales we are indeed averaging emission from several dense clouds. 
This is further supported by the typical emission found in these regions. 
In the selected beam-size regions (see Fig.~\ref{fig:Gallery}), we find typical integrated \CO intensities of $I_\mathrm{CO}\sim1 \times10^2\,$K\,km/s. 
A typical large molecular cloud with cloud masses of $M_\mathrm{cloud}\sim 1\times 10^5\,M_\odot$ would at our resolution correspond to integrated intensities of 1.3\,K\,km/s using a standard CO-to-H$_2$ conversion factor of $\alpha_\mathrm{CO}= 4.35\,M_\odot\,\mathrm{pc}^{-2}\,\left(\mathrm{K\,km/s} \right)^{-1} $.  
Therefore we can confirm that at our resolution we are likely averaging emission from several clouds.

\section{The \nnhp-to-\CO and \hcn-to-\CO relation}
\label{app:Additionalspikes}

\begin{figure*}[h]
    \centering
    \includegraphics[width = \textwidth]{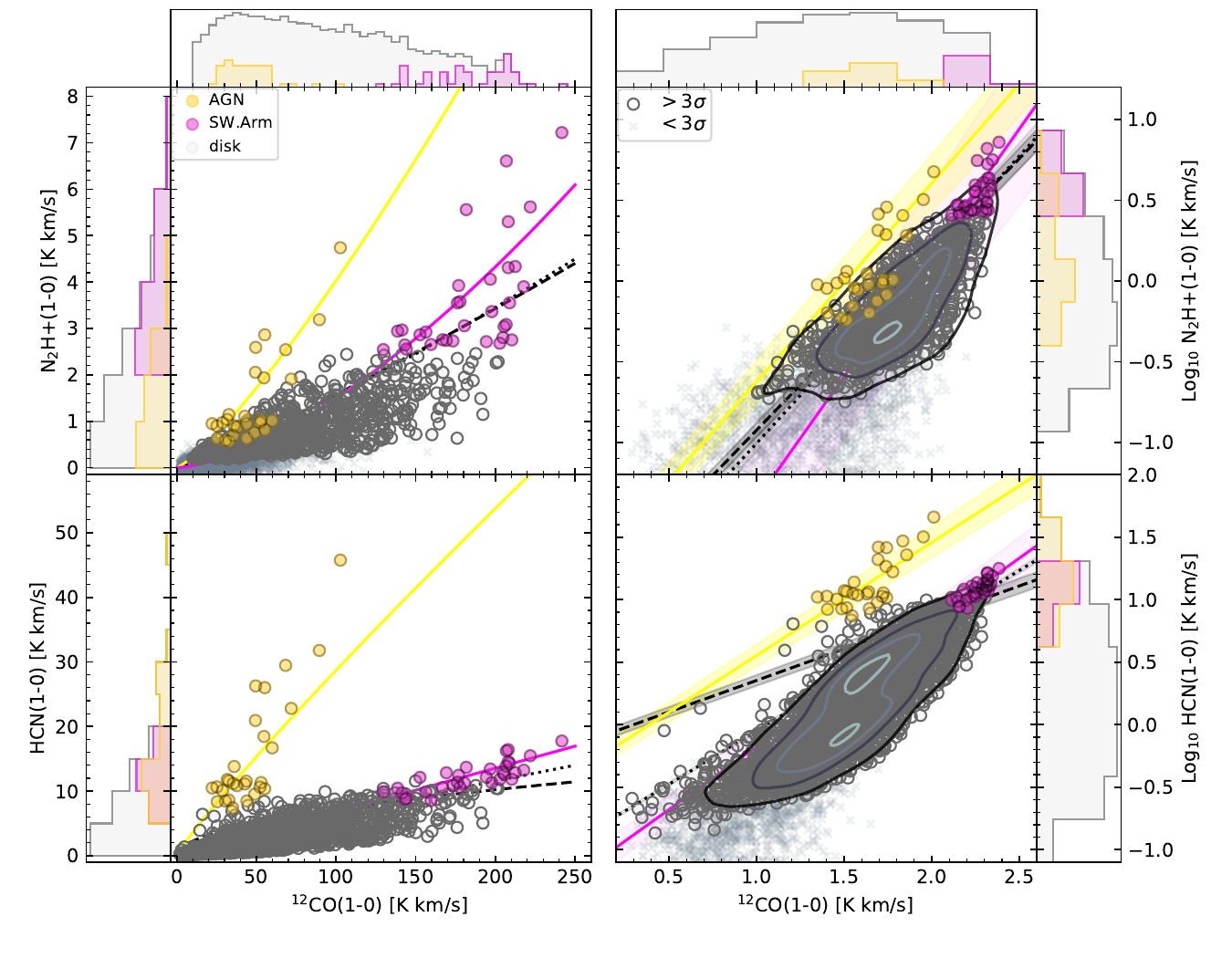}
    \caption{Pixel-by-pixel distribution of integrated \nnhp (top panels) and \hcn (bottom panels) as function of \CO emission, similar to Fig.~\ref{fig:Spikes} (a) in linear (left panels) and logarithmic (right panels) scaling. Subsets of pixels isolated in Fig.~\ref{fig:Spikes} are marked accordingly (pink: subset (c) SW.Arm, yellow: Subset (b) AGN). Power-law fits to the full data (black dashed), the data without the central points (black dotted) and the subsets are added (colors respectively). Fit uncertainties are only shown in log space to ease visibility.}
    \label{fig:Additionalspikes}
\end{figure*}

In addition to the \nnhp-to-\hcn distribution (Fig.~\ref{fig:Spikes}), we show the \nnhp-to-\CO and \hcn-to-\CO distributions in Fig.~\ref{fig:Additionalspikes}. 
We mark the same subset of pixels (AGN, SW.arm) identified in the \nnhp-to-\hcn distribution accordingly. 
We fit power-laws to all the data (dashed line), the data without the central points (dotted line), as well as the subsets of data (subset (b) AGN, subset (c) SW.arm). 
Fit parameters can be found in Table~\ref{tab:FitparamsAppendix} and more details on the fitting process are provided in Appendix~\ref{app:Fitting}.
Similarly to our previous findings, we see that the central data points follow a steeper trend than the rest of the data points. 
While the central points in the \hcn-to-\CO distribution do not overlap with the rest of the data, the central \nnhp-to-\CO distribution mostly overlaps with the rest of the data points and the central \nnhp-to-\CO trend is only slightly steeper than the global trend.  

The \nnhp-bright data points from the southern spiral arm (subset (c): pink) are also the brightest pixels in \CO.  
These points constitute a super-linear relation in the \nnhp-to-\CO plane. 
This is less clear for the \hcn-to-\CO distribution. 
While all fits to the \nnhp-to-\CO data and subsets are super-linear ($a>1$), all fits to the \hcn-to-\CO data and subsets are sub-linear ($a<1$) except for the fit to subset (c). 
We note, however, that there are additional data points in the \hcn-to-\CO distribution, which are elevated above the bulk distribution that might instead belong to the central subset (b), but are not selected since our selection of these subsets is visually determined based on the \nnhp-to-\hcn distribution.
While the fit to all data points shows a large increase in offset due to the central data points, the fit without the central subset might still be elevated due to the likely imperfect selection of data points impacted by the AGN. 

The fit to the central subset (b) of the \hcn-to-\CO distribution significantly deviates from the fit to the disk data without this subset, similar to our findings for the \nnhp-to-\hcn distribution. 
Unsuprisingly, this is not the case for the central fit of the \nnhp-to-\CO distribution, which agrees well with the fit to the disk data, as most of the central subset (b) overlaps with the disk data. 
This indicates that the mechanism driving the offset in line emission affects \nnhp less than \hcn (see discussion in Section~\ref{sec:Discussion}).

\setlength{\tabcolsep}{4pt}
\begin{table}
\begin{small}

\caption{Power-law fit parameters of \nnhp as a function of \CO, and \hcn as a function of \CO}\label{tab:FitparamsAppendix}
\centering
\begin{tabular}{l|llll}
\hline\hline
\noalign{\smallskip}
    Region & Power $a$ & Offset $b$ & $\rho_\mathrm{Sp}$ & $p$-value\\\noalign{\smallskip}
    \hline
\noalign{\smallskip}
        \multicolumn{3}{l}{\textbf{\nnhp-to-\CO}}\\
\noalign{\smallskip}
        All data & $1.12 \pm 0.02$ &$-2.05 \pm 0.04$ & $0.692 \pm 0.007$ & <0.001 \\
        All w/o AGN & $1.19 \pm 0.03$ & $-2.19 \pm 0.05$ & $0.751 \pm 0.005$ & <0.001\\
        AGN, yellow & $1.23 \pm 0.07$ & $-1.86 \pm 0.12$ & $0.69 \pm 0.03$ & <0.001\\
        SW.Arm, pink & $1.53 \pm 0.14$ & $-2.89 \pm 0.32$ & $0.82 \pm 0.02$ & <0.001\\ 
\hline
\noalign{\smallskip}
        \multicolumn{3}{l}{\textbf{\hcn-to-\CO}} \\
        All data & $0.50 \pm 0.02$ & $ -0.15 \pm 0.04$ & $0.836 \pm 0.003$ & < 0.001\\
        All w/o AGN & $0.86 \pm 0.01$ &$-0.90 \pm 0.01$ & $0.844 \pm 0.003$ & <0.001\\ 
        AGN, yellow & $0.91 \pm 0.05$ & $-0.35 \pm 0.08$ & $0.68 \pm 0.03$  & <0.001\\
        SW.Arm, pink & $1.01 \pm 0.05$ & $-1.18 \pm 0.11$ & $0.82\pm0.02$ & <0.001\\
        
\noalign{\smallskip}
\end{tabular}
\end{small}
\tablefoot{Fit parameters and spearman correlation coefficients according to a linear fit in log-log space similar to Table~\ref{tab:Fitparams} but for the \nnhp-to-\CO and \hcn-to-\CO distribution.}
\end{table}

\section{Fitting the \nnhp-to-\hcn, \nnhp-to-\CO and \hcn-to-\CO distribution}
\label{app:Fitting}

We fit all data points, the data subsets (b,c) as well as all disk data (all data without subset (b) AGN) of the \nnhp-to-\hcn, \nnhp-to-\CO and \hcn-to-\CO distributions with linear functions in logarithmic scaling. 
We only consider pixels with significant emission  ($>3\sigma$).
Parameters are fitted with \texttt{curve-fit} \citep[python scipy-optimize tool;][]{virtanen_scipy_2020}. 
We fit a linear function of shape $f(x)=ax+b$ to the logarithmic data with slope $a$ and offset $b$. Following error propagation, the corresponding uncertainty at each x value is $\Delta f(x) = \sqrt{\left(\Delta a x \right)^2 +  \left(\Delta b\right)^2}$, with uncertainties $\Delta a$, $\Delta b$ accordingly. The discrepancy to the literature fit is measured as $\sigma = |f - f_{lit}|/\sqrt{\Delta f^2 + \Delta f_{lit}^2}$, which is dependent on x. We provide the average discrepancy in the range over which our data are measured.

Uncertainties are estimated by perturbing each pixel by a random Gaussian value with standard deviation at the corresponding noise value, and randomly jackknifing 10\% of the data before either calculating fit-parameters or Spearman correlation coefficients. Repeating this 100 times yields the standard deviation as uncertainty.

\section{Quantifying the scatter of line ratios}

We investigate regions with increased scatter, as well as quantify the scatter between \nnhp and \hcn, \hcn and \CO as well as \nnhp and \CO.

\subsection{Disentangling emission from the southern spiral arm}
\label{App:scatter}

\begin{figure}
    \centering
    \includegraphics[width = 0.5\textwidth]{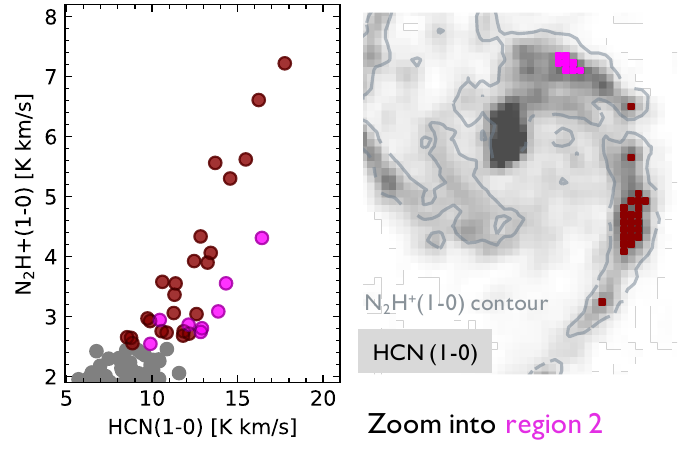}
    \caption{Close-up of \hcn and \nnhp emission (left panel) of \nnhp-bright pixels (subset (c), see Fig.~\ref{fig:Spikes}), as well as their location in the disk (right panel, compare to panel b of Fig.~\ref{fig:Spikes}). We separate the pixels into two sub-regions (red, pink)}
    \label{fig:ZoomSpike}
\end{figure}

In Sec.~\ref{sec:N2HP_HCN_relation} we isolated pixels that are bright in \hcn and \nnhp. 
The region with pixels brightest in \nnhp (subset (c): pink points in Fig.~\ref{fig:Spikes}) shows a comparably large scatter in \nnhp emission, with the emission varying by nearly a factor of 2 at similar levels of \hcn flux (at $I_\mathrm{HCN} \sim 18\,$K~km/s we find $I_\mathrm{N_2H^+} \sim 4-7.5\,$K~km/s).
A closer look (Fig.~\ref{fig:ZoomSpike}) reveals that this emission originates from different spatial locations in the disk, one with a shallower \nnhp-to-\hcn distribution located in the north-western part of the molecular ring (pink circles), the other with a steeper \nnhp-to-\hcn distribution from the south-western part of the same spiral arm (dark red circles). 

As noted in Sec.~\ref{sec:N2HP_HCN_relation}, the data points from the southern spiral arm (red points in Fig.~\ref{fig:ZoomSpike}) drive the non-linear but logarithmic relation between \nnhp and \hcn emission. 
Interestingly, the northern region is located close to the AGN jet major axis \citep{querejeta_agn_2016}, and might be potentially impacted by the AGN as well, though the number of data points is lower. 

\subsection{Measuring the scatter of line ratios}
\label{App:scatter2}
We quantify the scatter of the \nnhp-to-\hcn, \hcn-to-\CO and the \nnhp-to-\hcn distribution as follows: 
For this analysis, we exclude the central subset (b) that exhibits a quite different distribution from the rest of the data, as well as data points below the $3\sigma$ noise level. 
We subtract the corresponding best-fit value (using the fit parameters when excluding the center) from each data point in the according distribution. 
The obtained data has an average scatter of 
$\sim0.14$ dex for \nnhp as a function of \hcn emission, $0.19$ dex for \nnhp as a function of \CO emission and $0.29$ dex for \hcn as a function of \CO emission. 

We conclude the following: 
a) While all lines span a range of $\gtrsim 1.5$ dex in intensity, we find $\sim10\%$ scatter, indicating that all lines are well correlated. 
b) The scatter of \nnhp as a function of \hcn is least, indicating a tighter correlation between \nnhp and \hcn than any of those lines with \CO. 
Since these results are strongly dependent on the fit and thus the fitting tool used, we suggest that the reader should take these results with caution.

\end{appendix}
\end{document}